\documentclass[aps,pre,onecolumn,amsmath,superscriptaddress]{revtex4}
\usepackage{graphicx}
\usepackage{bm,color,textcomp}
\usepackage{amsmath,amsfonts,amssymb}

\usepackage{subfig}%per mettere le figure affiancate
\usepackage{graphicx}

\usepackage[normalem]{ulem}

\begin{document}
%%%%%%%%%%%%%%%%%%%%%%%%%%%%%%%%%%%%%%%%%%

\title{Spatial velocity correlations in inertial systems of Active Brownian Particles}

% Authors, for the paper (add full first names)
\author{Lorenzo Caprini} 
\affiliation{School of Sciences and Technology, University of Camerino, Via Madonna delle Carceri, I-62032, Camerino, Italy.}
\author{Umberto Marini Bettolo Marconi}
\affiliation{School of Sciences and Technology, University of Camerino, Via Madonna delle Carceri, I-62032, Camerino, Italy.}

\date{\today}

\begin{abstract}
Recently, it has been discovered that systems of Active Brownian particles (APB) at high density organise their velocities into coherent domains showing large spatial structures in the velocity field. Such a collective behavior occurs spontaneously, i.e. is not caused by any specific interparticle force favoring the alignment of the velocities. This phenomenon was investigated in the absence of thermal noise and in the overdamped regime where inertial forces could be neglected. In this work, we demonstrate through numerical simulations and theoretical analysis that the velocity alignment is a robust property of ABP and persists even in the presence of inertial forces and thermal fluctuations. We also show that a single dimensionless parameter, such as the P\'eclet number customarily employed in the description of self-propelled particles, is not sufficient to fully characterize such a phenomenon neither in the regimes of large viscosity nor small mass. Indeed, the size of the velocity domains,  measured through the correlation length of the spatial velocity correlation, remains constant when the swim velocity increases while decreases as the rotational diffusion becomes larger. We find that the spatial velocity correlation depends on the inertia but, contrary to common belief, are non-symmetrically affected by mass and inverse viscosity variations. We conclude that in self-propelled systems, at variance with passive systems, variations of the inertial time (mass over solvent viscosity) and mass act as independent control parameters. Finally, we highlight the non-thermal nature of the spatial velocity correlations that are fairly insensitive both to solvent and active temperatures.
\end{abstract}

\newcommand{\betadef}{\frac{1}{\tau}}
\newcommand{\alphadef}{\frac{\omega_q^2}{\gamma}}
\newcommand{\br}{{\bf r}}
\newcommand{\bu}{{\bf u}}
\newcommand{\bR}{{\bf x}}
\newcommand{\bRz}{{\bf x}^0}
\newcommand{\bk}{{ \bf k}}
\newcommand{\bx}{{ \bf x}}
\newcommand{\vv}{{\bf v}}
\newcommand{\nb}{{\bf n}}
\newcommand{\mb}{{\bf m}}
\newcommand{\bq}{{\bf q}}
\newcommand{\rb}{{\bar r}}

\newcommand{\eeta}{\boldsymbol{\eta}}
\newcommand{\xxi}{\boldsymbol{\xi}}

\maketitle

%%%%%%%%%%%%%%%%%%%%%%%%%%%%%%%%%%%%%%%%%%

%%%%%%%%%%%%%%%%%%%%%%%%%%%%%%%%%%%%%%%%%%
\section{Introduction}

Many systems of biological or technological interest
display fascinating spatial velocity correlations extending over lengths larger than the size of the individual constituents.
This phenomenon is an example of the intriguing non-equilibrium behavior typical of active~\cite{marchetti2013hydrodynamics, elgeti2015physics, gompper20202020} 
and granular matter systems~\cite{van1999randomly,baldassarri2002cooling} 
and is in stark contrast with the 
observed behavior characteristic of equilibrium colloidal  suspensions 
where  the particle velocities are uncorrelated and follow the Maxwell-Boltzmann distribution.

Colonies of bacteria, such as Bacillus subtilis or Myxococcus xanthus, display spatial velocity correlations  exponentially  decaying with a correlation length much larger than the typical bacterium size~\cite{dombrowski2004self, peruani2012collective, wioland2016ferromagnetic}. The velocity field of bacteria forms vortex-domains or clusters where the velocities are mutually aligned and 
continuously rearrange according to different patterns.
This phenomenon occurs at large densities and is often called bacterial turbulence and has been mostly investigated in the framework of hydrodynamic phenomenological theories~\cite{wensink2012meso, dunkel2013fluid, urzay2017multi, james2018turbulence}.
Particle-based numerical studies have reproduced the formation of velocity domains either in models containing an explicit velocity alignment interaction term~\cite{grossmann2014vortex} or in models where the observed rich variety of polar phases~\cite{grossmann2020particle} was mainly due to the elongated shape typical of many  species of bacteria.

More recently, the experimental study of cell monolayers has revealed similar spatial structures in the velocity field 
extending over a range of $\sim10-20$ microns for mesenchymal cell up to $\sim 500$ microns for very adhesive epithelial cells reaching also $\sim 50$ times the typical size of the single cell \cite{petitjean2010velocity}.
Many cells, such as the typical Madin-Darby Canine Kidney (MDCK) cells~\cite{Heinrich2020size} or human bronchial epithelial cells (HBEC)~\cite{blanch2018turbulent}, are not elongated 
%having just a certain degree of deformability, %\cite{saw2017topological}
but still form large groups with correlated velocities often organizing in vortex structures~\cite{blanch2018turbulent, henkes2020dense} (without showing the formation of polar bands) and give rise to velocity correlations exponentially decaying in space ~\cite{garcia2015physics, basan2013alignment}.
To explain these behaviors, several models have been proposed~\cite{alert2020physical}.
At the particle level, alignment interactions between particle polarizations or particle velocities have been often included in the cell dynamics~\cite{sepulveda2013collective, sarkar2020minimal}. However, in recent studies, these phenomenological interactions have been replaced by additional frictional forces~\cite{garcia2015physics} or complex anti-alignment interactions of biological origin~\cite{smeets2016emergent} that could also give rise to a similar phenomenology.

%These systems form domains where the velocities are aligned without showing the formation of polar bands typically observed in Vicsek-like models.
Despite their different origins, the common feature of these systems is the formation of domains with correlated velocities even in the absence of the polar bands that instead are typically observed in Vicsek-like models.
At variance with the mentioned theoretical approaches, the local velocity alignment %with vanishing global polar order 
has been recently reproduced via dissipative stochastic dynamics without introducing any explicit alignment interactions between the particle orientations~\cite{caprini2020spontaneous, caprini2020hidden, caprini2020time} or some kind of local interaction between particle velocity and self-propulsion.
Dense systems of purely repulsive Active Brownian Particles (ABP) form domains where the velocities are aligned or arranged in vortex-like patterns when they attain hexatic or solid order~\cite{caprini2020hidden} or in the dense phase of the non-equilibrium phase-coexistence~\cite{caprini2020spontaneous}, known as Motility induced phase separation (MIPS)~\cite{cates2015motility, gonnella2015motility, bialke2015active}.
The ABP already contains the following minimal ingredients producing velocity patterns: i) persistent self-propulsion forces and ii) purely repulsive interactions.
However, so far these results have been obtained through theoretical analysis and simulations neglecting
two important aspects: the inertial forces and thermal noise due to the molecules of the solvent.
In apparent contradiction with the results of Refs.~\cite{caprini2020spontaneous, caprini2020hidden, caprini2020time},
a successive investigation, based on thermal overdamped ABP, ~\cite{caporusso2020motility} and focused on micro-phase motility induced phase separation did not reveal the presence of spatial velocity correlations.
Two natural questions arise: i) does the velocity alignment in ABP systems  occur only in the absence of thermal fluctuations? ii) Is this ordering  suppressed if one takes into account the effect of the acceleration?

We anticipate the main result of the present study:
the spatial patterns in the velocity field of active systems survive in the case of underdamped active dynamics and thermal noise.
Our investigation also proves three important results derived  by combining numerical and theoretical methods:
\begin{itemize}
\item[i)] The inadequacy of the so-called P\'eclet number, as a single active force dimensionless parameter, to understand the dynamical collective phenomena.
Indeed, we unveil the non-symmetric role of persistence time and swim velocity, being the spatial velocity correlation function independent of the latter but deeply affected by the former.
\item[ii)] Asymmetric role of mass and inverse viscosity in the velocity correlations functions whose changes are not controlled only by the inertial time (mass over viscosity), but depend on both parameters. 
\item[iii)] Marginal role of thermal and active temperatures for the dynamical collective phenomena presented so far.  The temperature increase does not affect the correlation length of the spatial velocity correlation, revealing a dynamical scenario fairly different from what one expects for equilibrium ferromagnetic systems.
\end{itemize}

The article is structured as follows: in Sec.~\ref{Sec:Model}, we introduce the model describing the self-propelled system in the underdamped regime and, in Sec.~\ref{sec:3}, we present the velocity alignment phenomenology.
Secs.~\ref{sec:4},~\ref{sec:5} and~\ref{sec:6} discuss the role of the active force, inertial forces and temperature.
Finally, we conclude by summarizing the main results and presenting some final remarks. 

%%%%%%%%%%%%%%%%%%%%%%%%%%%%%%%%%%%%%%%%%%%%%%%%%%%%%%%%%%%%%%%%%%%%%%%%
\section{Model}
\label{Sec:Model}
%%%%%%%%%%%%%%%%%%%%%%%%%%%%%%%%%%%%%%%%%%%%%%%%%%%%%%%%%%%%%%%%%%%%%%%%

In order to investigate the collective dynamics of a system of inertial self-propelled particles, we perform numerical simulations of the underdamped version of the ABP model, while to build a theoretical framework, we employ the Active Ornstein-Uhlenbeck (AOUP) model containing  the same deterministic force terms. In the two models, the active forces  are different but share
 similar statistical properties.
We resort to this procedure because it greatly simplifies the theoretical analysis.
Both the AOUP and the ABP have been successfully employed to reproduce many aspects of the active matter phenomenology including accumulation near an obstacle, velocity correlations and entropy production~\cite{caprini2018active,marconi2016velocity,fodor2016far,marconi2017heat,das2018confined, caprini2019activechiral, maggi2020universality}.
 The underdamped ABP equation of motion, describing a system of interacting self-propelled particles of mass $m$, are:
\begin{subequations}
\label{eq:motion}
\begin{align}%\label{eq:motion}
%\begin{aligned}
\dot{\mathbf{x}}_i&=\mathbf{v}_i \,,\\
m\dot{\mathbf{v}}_i& = - \gamma \mathbf{v}_i +\mathbf{F}_i + \mathbf{f}^a_i + \sqrt{2 \gamma  T} \,\boldsymbol{\eta}_i \,,
%\end{aligned}
\end{align}
\end{subequations}
where $\mathbf{x}_i$ and $\mathbf{v}_i$ represent the particle position and velocity, respectively.
The drag coefficient, $\gamma$, and the solvent temperature, $T$, determine the thermal diffusion coefficient, $D_t$
via the Einstein relation, $\gamma D_t= T/m$.
The term $\boldsymbol{\eta}$ is a white noise vector with zero average and unit variance accounting for the random collisions between the self-propelled particle and the particles of the solvent, such that $\langle \boldsymbol{\eta}_i(t) \boldsymbol{\eta}_j(t')\rangle=\boldsymbol{\delta}(t-t')\delta_{ij}$.
As for equilibrium colloids, the solvent exerts a Stokes drag force proportional to $\mathbf{v}_i$.
Often, the thermal diffusivity of active colloidal and bacterial suspensions~\cite{bechinger2016active}, is negligible compared to the effective diffusivity produced by the active force.
 The effect of inertia is also considered not to be important in the case of typical active particles such as microscopic self-propelled colloids or bacteria swimming in solution.
However, this approach needs to be reconsidered in the light of recent studies focused on the interplay between inertia and active forces~\cite{lowen2020inertial, mandal2019motility, caprini2020inertial, petrelli2020effective, dai2020phase, su2020inertial, vuijk2020lorentz} motivated by the existence of experimental macroscopic systems, such as vibro-robots~\cite{scholz2018inertial, dauchot2019dynamics} or camphor surfers~\cite{leoni2020surfing} which behave as active particles.

The particle interactions are represented by the force $\mathbf{F}_i = - \nabla_i U_{tot}$, where $U_{tot} = \sum_{i<j} U(|\mathbf{x}_i -\mathbf{x}_j|)$ is a pairwise potential. 
We choose $U$ as a shifted and truncated Lennard-Jones potential 
~\cite{redner2013structure, caprini2020hidden}:
\begin{equation}
\label{eq:interactionpotential}
U(r) = 4\epsilon \left( \left(\frac{\sigma}{r}\right)^{12}- \left(\frac{\sigma}{r}\right)^6  \right) \,,
\end{equation}
for $r\leq 2^{1/6}\sigma$ and zero otherwise.
The constants $\epsilon$ and $\sigma$ determine the energy unit and the nominal particle diameter, respectively.
In the spirit of minimal modeling, the self-propulsion is represented through a stochastic force, namely $\mathbf{f}_i^a$.
At this level of description, the details about the chemical or mechanical origin of the self-propulsion~\cite{bechinger2016active, shaebani2020computational, gompper20202020, marchetti2013hydrodynamics} are not specified.
This force drives the system far from equilibrium~\cite{fodor2016far, dabelow2019irreversibility} and determines a persistent motion in a random direction lasting for a time smaller than a characteristic  persistence time, $\tau$.
The two dimensional ABP self-propulsion is a force with constant modulus $f_0$ and time-dependent orientation $\mathbf{n}_i=(\cos{\theta_i}, \sin{\theta_i})$:
\begin{equation}
\label{eq:activeforce}
\mathbf{f}^a_i=f_0 \mathbf{n}_i \, .
\end{equation}
The angle $\theta_i$ performs a Brownian motion:
\begin{equation}
\label{eq:theta}
\dot{\theta}_i = \sqrt{2D_r} \chi_i \,,
\end{equation}
being $\chi_i$  a white noise with zero average and unit variance and $D_r=1/\tau$ a rotational diffusion coefficient determining how persistent is the propagation direction.
The parameter $f_0$ fixes the swim velocity induced by the self-propulsion: 
\begin{equation}
\label{eq:def_swimvelocity}
v_0=\frac{f_0}{\gamma} \,.
\end{equation}
%The larger the friction the larger is the value of $f_0$ needed to induce a given value of the swim velocity $v_0$.
Finally, we introduce the active temperature:
\begin{equation}
\label{eq:def_activetemperature}
T_a = f_0^2 \frac{\tau}{\gamma} = \gamma \tau v_0^2 \,.
\end{equation}
in agreement with previous definitions employed for overdamped active dynamics~\cite{berthier2019glassy, caprini2020inertial}.
%$T_{a}$ plays the same role as the equilibrium temperature $T$ in the expression for the long-time diffusion coefficient of an isolated particle~\cite{berthier2019glassy}. 
This parameter will play a relevant role in the following.

The AOUP model~\cite{berthier2017active, maggi2014generalized, woillez2020nonlocal, caprini2018activeescape, wittmann2016active, martin2020statistical, maggi2017memory, caprini2018linear, szamel2014self}, %wu2000particle, koumakis2014directed
employed to ease the theoretical analysis replaces the ABP self-propulsion~\eqref{eq:activeforce} by an Ornstein-Uhlenbeck process:
\begin{equation}\label{eq:motion2}
\tau\dot{\mathbf{f}}^a_i = -\mathbf{ f}^a_i + f_0\sqrt{2 \tau} \boldsymbol{\xi}_i \,,
\end{equation}
where $\boldsymbol{\xi}_i$ is a white noise vector with zero average and unit variance, such that $\langle \boldsymbol{\xi}_i(t)  \boldsymbol{\xi}_j(s) \rangle = \delta_{ij} \delta(t-s)$.
In the AOUP, the modulus of $\mathbf{f}^a$ is not held rigidly fixed but fluctuates around the mean value $f_0$.
The correlation time, $\tau$, of the active force, are chosen to have a common value in AOUP and ABP~\cite{farage2015effective, caprini2019comparative}.  In both models
the self-correlation of the active force decays in time with an exponential law.

Regarding the aptness of the AOUP for adequately reproducing the salient features of the ABP, we mention a recent study~\cite{caprini2020active} of the single-particle velocity distribution in the case of dense active solid configurations, similar to those analyzed in this paper. In that work, we concluded that in the large persistence regime (i.e. for a broad range of $\tau$ including the values analyzed in this work) the ABP single-velocity properties are well-described by those of an AOUP system at variance with the small persistence regime.

\section{Velocity alignment}\label{sec:3}

%------------------------ FIG.1--------------------------------------
\begin{figure*}[!t]
\centering
\includegraphics[width=0.99\linewidth,keepaspectratio]{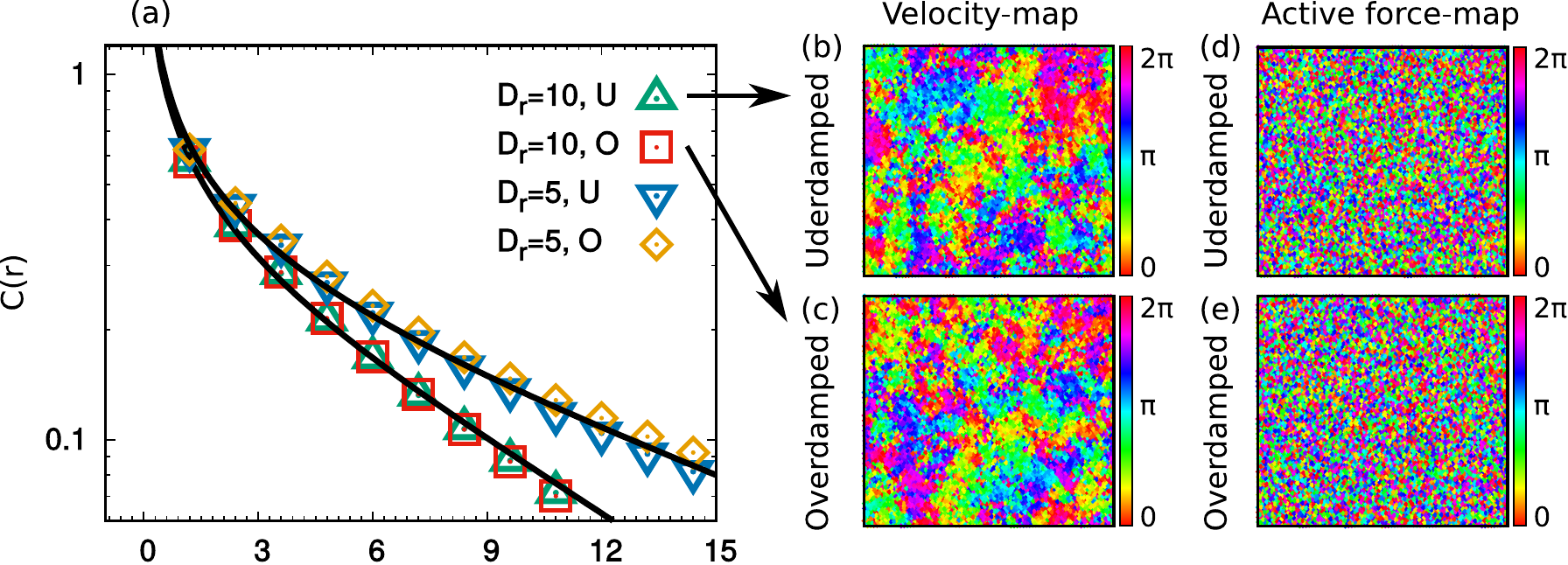}
\caption{\label{fig:corr_self}
Comparison between the velocity domains of overdamped and underdamped dynamics. Panel (a): spatial velocity correlation, $C(r)=\langle \mathbf{v}(r) \cdot \mathbf{v}(0) \rangle/\langle \mathbf{v}^2 \rangle$, for two different values of $D_r$, as detailed in the legend.  
For both values, we compare the correlation obtained  via underdamped dynamics Eq.~\eqref{eq:motion} (denoted by the symbol U) 
with the one corresponding to overdamped dynamics  Eq.~\eqref{eq:motion_over} (symbol O).
The dashed black lines represent the theoretical predictions, obtained by fitting the functional form given by Eq.~\eqref{eq:vv_realspace} with the function $f(r)=a\,e^{-r/\lambda}/r^{1/2}$, where $\lambda$ is given by Eq.~\eqref{eq:lambda} and $a$ and $b$ are two positive fitting parameters.
Panel (b),(c),(d) and (e):  Snapshot configurations for $D_r=10$ relative to underdamped dynamics (panels (b) and (d)) and to overdamped dynamics (panels (c) and (e)). Particles are colored according to the velocity direction in panels (b), (c) and according the orientational angle, $\theta$ (
identifying the direction of the active force),  in panels (d), (e), respectively.
The velocity vector in the overdamped case are represented by $\dot{\mathbf{x}}$ as described in Appendix~\ref{app:1}.
%$\bar{x}$ has directly measured numerically and reads $\bar{x}= 0.91$.
The simulations have been obtained using $\gamma/m=10^2$, $\epsilon=10^2$, $T=10^{-1}$, $f_0=5\times 10^3$, corresponding to a swim velocity of $v_0=50$.
}
\end{figure*}
%---------------------------------------------------------------------

We have integrated numerically the equations~\eqref{eq:motion} and~\eqref{eq:theta} for a system of $N$ particles moving in a square domain of size $L$ with periodic boundary conditions. 
The simulations are performed keeping fixed the packing fraction $\phi=N/L^2 \sigma^2\pi/4$ in such a way that the system attains a solid configuration without showing changes in the positional structure of the system for a broad range of activity parameters (both $f_0$ and $D_r$). 
Indeed, it is known that
%, for a broad range of packing fractions, 
the increase of both $f_0$ (or equivalently of $v_0$) and $\tau$ induces the solid-hexatic and finally the hexatic-liquid transition~\cite{bialke2012crystallization, digregorio2018full, caprini2020hidden}.
A further increase of $f_0$ and $\tau$ leads to a non-equilibrium phase-coexistence that, at variance with passive Brownian particles, occurs even in the absence of attractive interactions~\cite{fily2012athermal, cates2013active, buttinoni2013dynamical, stenhammar2015activity, solon2015pressure, mallory2018active, shi2020self}.  
This phenomenon, known as motility induced phase separation (MIPS) %~\cite{cates2015motility, gonnella2015motility, bialke2015active} ,
is due to the particle slowdown caused by interactions~\cite{redner2013structure}. 

Refs.~\cite{caprini2020spontaneous,caprini2020hidden} (for a phase-separated and homogeneous liquid, hexatic and solid configurations, respectively) demonstrated the spontaneous occurrence of velocity alignment in the case of athermal ABP in the overdamped regime despite the absence of any form of alignment interaction. 
As a first result, we show that the spontaneous velocity alignment occurs even in the case of the 
underdamped dynamics modeled by Eqs.~\eqref{eq:motion}, that account for both the 
finite particle acceleration and thermal fluctuations induced by the solvent. 
Fig.~\ref{fig:corr_self}, represents a pair of snapshots illustrating the comparison between a system governed by Eqs.~\eqref{eq:motion} with $\gamma=10^2$ and $m=1$ and a system evolving with the overdamped dynamics whose details are reported in Appendix \ref{app:1}.
In particular, in panels (b) and (c), the color-map represents the velocity direction of each particle while, in panels~(d) and~(e), the orientation of the self-propulsion. In the former case, the particles are colored according to the angle formed by the velocity $\mathbf{v}_i$ 
of each particle
with the $x$ axis, while, in the latter case, according to the angle $\theta_i$ of the self-propulsion.
While the self-propulsion directions are random without showing any spatial structure (as expected from Eq.~\eqref{eq:theta}), large domains containing aligned velocities are observed.
It means that $\mathbf{v}_i$ does not coincide with $\mathbf{f}^a_i$ in dense configurations where the interparticle interactions are not rare events.
The same scenario could be detected in the bulk of the dense phase of MIPS that reaches very large packing fractions attaining configurations that could even display the hexatic or almost-solid orders \cite{caprini2020spontaneous}.

To quantify the size of the velocity domains we study the spatial velocity correlation function, $C(r)$, defined as:
$$
C(r)=\frac{\langle \mathbf{v}(r) \cdot \mathbf{v}(0) \rangle}{\langle \mathbf{v}^2 \rangle} \,,
$$ 
normalized by dividing by the velocity variance, $\langle \mathbf{v}^2 \rangle$. The associated correlation length provides a measure of the average size of a velocity domain since particles not belonging to the same domain display uncorrelated velocities.
The observable $C(r)$ is reported in Fig.~\ref{fig:corr_self}~(a) for two different values of $D_r$ both for the underdamped and the overdamped dynamics for large values of $\gamma$ such that the inertial forces play a marginal role. Two values of $D_r$ are reported, such that $\tau=1/D_r \gg m/\gamma$, and both reveal a fair agreement between overdamped and underdamped dynamics.
As already shown in Ref.~\cite{caprini2020hidden}, the spatial velocity correlation decreases slower as $D_r$ is increased and, in particular, the correlation length scales as $\tau^{1/2}$ with $\tau=1/D_r$ in the overdamped regime. 
How that scaling with $\tau$ would be modified due to inertial effects is described in Sec.~\ref{sec:5}.

\subsection{Theoretical prediction}

We have extended to the dynamics~\eqref{eq:motion}
the analytical method previously employed in the study of the spatial velocity correlation functions in the case of overdamped ABP in dense configurations ~\cite{caprini2020spontaneous, caprini2020hidden, caprini2020time}.
  The details of the calculations are reported in Appendix~\ref{app:2} and lead to the following formula
for the  Fourier transform of the steady-state  equal-time velocity correlation:
\begin{equation}
\label{eq:vv_fourierspace}
\langle \hat{\mathbf{v}}(\mathbf{q})\cdot \hat{\mathbf{v}}(-\mathbf{q})\rangle = \frac{2T}{ m}  +\frac{2 T_a }{m}\frac{1}{1+\tau/\tau_I} \frac{1}{1+\frac{\tau^2}{1+\tau/\tau_I}\omega^2(\mathbf{q})}
\end{equation}
where $\tau_I=m/\gamma$ is the inertial time and $T_a$ the active temperature, defined in Eq.~\eqref{eq:def_activetemperature}.
The vector $\mathbf{q}$ is a vector of the Fourier space and $\hat{\mathbf{v}}(\mathbf{q})$ is the Fourier transform of the velocity vector. % we have reported the final expression in the continuum notation for presentation reasons.
The frequency $\omega(\mathbf{q})$ in the long-wavelength limit, $\mathbf{q}\to 0$, reduces to:
\begin{equation}
\label{eq:omega_fourier}
\omega^2(\mathbf{q}) \approx  \frac{3  \omega_E^2}{2 }  \bar x^2 \mathbf{q}^2 \,
\end{equation}
with
$$
\omega_E^2 = \frac{1}{2 m}\left(U''(\bar x) + \frac{U'(\bar x)}{\bar x} \right) \, .
$$
The terms $U'(\bar{x})$ and $U''(\bar{x})$ represent the first and the second derivative of $U$ calculated at $\bar{x}$, the average distance between two nearest neighbor particles.
The full expression for $\omega^2(\mathbf{q})$ is reported in Appendix~\ref{app:2}.

Using formula~\eqref{eq:omega_fourier},
we can find (see Appendix~\ref{app:3}) the following expression
for the real space velocity correlation, holding for large distances (at least, $r>\sigma$):
\begin{equation}
\label{eq:vv_realspace}
C(r) 
 \approx \frac{2}{\langle \mathbf{v}^2\rangle}\frac{T_a }{m} \frac{1}{1+\tau/\tau_I} \frac{\bar{x}^2}{\lambda^2} \left( \frac{\lambda}{8 \pi r} \right)^{1/2} e^{-r/\lambda} \,,
\end{equation}
where the correlation length $\lambda$ is given by
\begin{equation}
\label{eq:lambda}
      \lambda^2  =   \frac{3}{2 }\bar x^2 \frac{  \omega_E^2 \tau^2}{1+\frac{\tau}{\tau_I}} \,.
\end{equation}
The overdamped result derived in Refs.~\cite{caprini2020spontaneous, caprini2020hidden, caprini2020time} is recovered in the limit $\tau_I \ll \tau$, i.e. when the solvent viscosity is sufficiently large (or the particle mass sufficiently small) compared to the persistence time of the active force.

For some choices of the parameters of the active force, it is possible to obtain large values of $\lambda$ so that a huge group of particles moves in the same direction.
Hence, to exclude undesired finite-size effects, we always performed simulations in such a way that the condition $L \gg \lambda$ is satisfied. Such a condition guarantees that the spatial velocity correlation approaches zero by avoiding finite-size effects and is fundamental to get results consistent with the theoretical analysis.
If this condition is not fulfilled, particles could form a single velocity domain (spanning the entire simulation box) oriented in a direction that changes with a typical time $\propto \tau$.
This state is known as active traveling crystals~\cite{menzel2013traveling, menzel2014active, briand2018spontaneously} and disappears performing simulations with larger boxes.

On the other hand,
 Eq.~\eqref{eq:vv_realspace} displays a non-physical divergence at the origin and does not correctly reproduce the behavior of $C(r)$ for small separations, namely $r < \sigma$. %, and, in particular the scaling of the velocity variance with the parameter of the active force.
The divergence is determined by the absence of an upper cutoff in the $\mathbf{q}$-integral
 that is used to  derive analytically the Fourier anti-transform of Eq.~\eqref{eq:vv_fourierspace}.
The divergence disappears by considering the correct integration limits when anti-transforming Eq.~\eqref{eq:vv_fourierspace}.  In Appendix~\ref{app:4}, we calculate the variance of the velocity distribution employing the exact expression of $\omega(\mathbf{q})$ and 
obtain the analytical expression of the kinetic temperature, $T_k =m \langle \mathbf{v}^2 \rangle/2$, in the presence of inertial forces and thermal noise:
%We calculate analytically the kinetic temperature, $T_k =m \langle \mathbf{v} \cdot \mathbf{v} \rangle/2$, 
%in the presence of inertial forces and thermal noise, and obtain:  
\begin{equation}
\label{eq:prediction_kinetictemperature2}
%T_k = T +   \frac{T_a}{1+\tau/\tau_I+6 \omega_E^2\tau^2} \frac{6 }{\pi z \sqrt{c}} {\bf K}
T_k = T +   \frac{T_a}{1+\tau/\tau_I+6 \omega_E^2\tau^2}  \frac{ \mathcal{I}}{\pi} \,,
\end{equation}
where the term $\mathcal{I}$ is a function of $\tau$, $\tau_I$ and $\omega_E$. 
The term $\mathcal{I}$ in Eq.~\eqref{eq:prediction_kinetictemperature2} is reported in Appendix~\ref{app:4} and contains the complete elliptic integral of the first kind.
Here, we just stress that $\mathcal{I}$ does not show any dependence on $T_a$ or $T$.
 Formula~\eqref{eq:prediction_kinetictemperature2} generalizes the overdamped result of  Ref.~\cite{caprini2020active}, (derived for overdamped ABP, such that $\tau_I \ll \tau$), and provides an analytical prediction for the kinetic temperature.   

We remark that the predictions regarding the spatial velocity correlations and kinetic temperature hold in the solid-like regime and, as already shown in Ref.~\cite{caprini2020hidden}, break down when the solid-hexatic transition takes place and the number of defects becomes statistically relevant.
In addition, expression~\eqref{eq:vv_realspace} can be used to extract $\lambda$ from simulations through numerical fits and compare it with the prediction~\eqref{eq:lambda}.

In the next sections, we report an extensive numerical study varying both the parameters of the active force and  inertial force taking advantage of the comparison with our theory.
The effect of the density increase has been already discussed in Ref.~\cite{caprini2020hidden} where the phase diagram (density, $\rho$, vs $\tau$ plotting $\lambda$ as a color gradient) has been reported. In this paper, we do not perform numerical investigation varying the density but recall  that the larger $\rho$ the larger $\lambda$.
In the solid-like phase, this is consistent with Eq.~\eqref{eq:lambda}, since the increase of $\rho$ produces the decrease of $\bar{x}$ and, thus, the increase of the factor $U''(\bar{x})+ U'(\bar{x})/\bar{x}$ appearing in the expression for $\omega^2_E$ that is proportional to $\lambda$.

\section{Role of the self-propulsion}\label{sec:4}

%------------------------ FIG.2--------------------------------------
\begin{figure}[!t]
\centering
\includegraphics[width=0.5\linewidth,keepaspectratio]{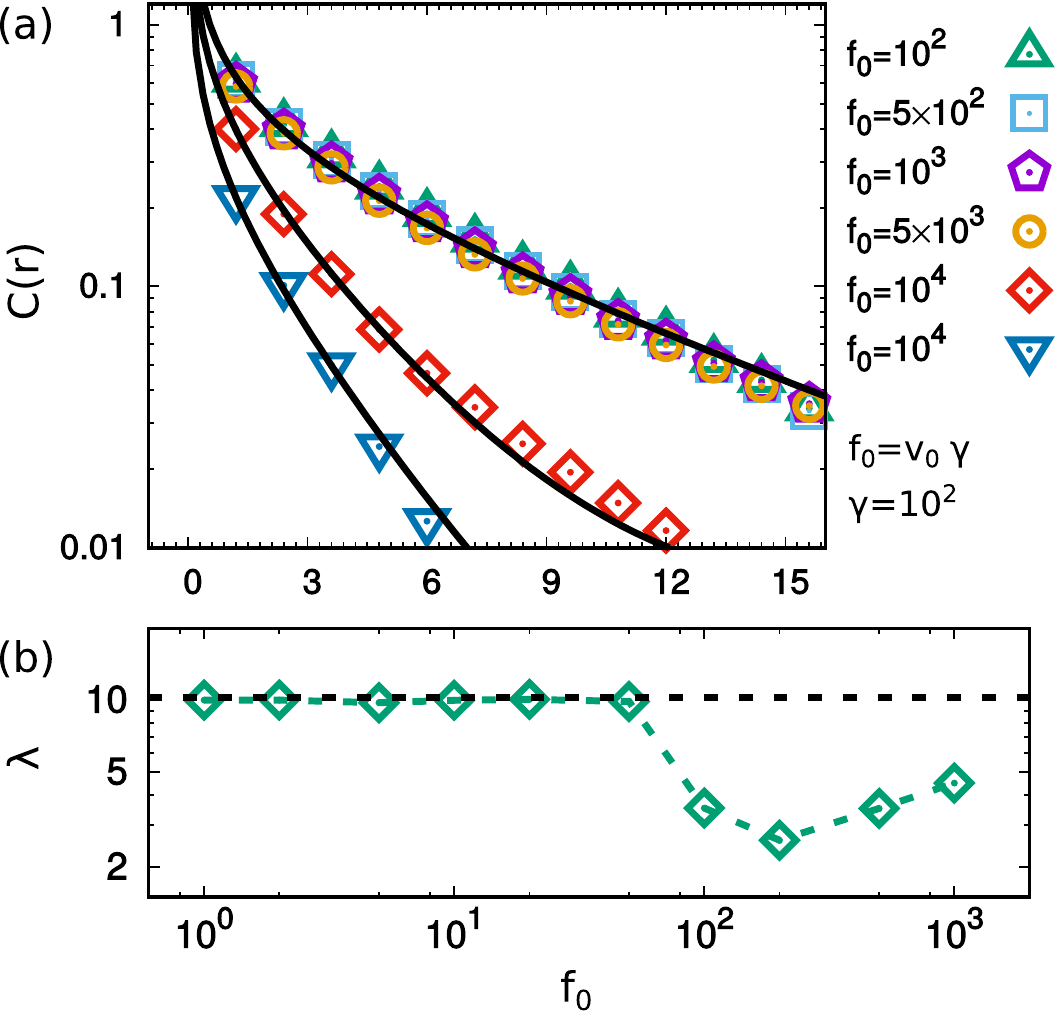}
\caption{\label{fig:corr_self}
Spatial velcity correlation as a function of the self-propulsion intensity. Panel (a): spatial velocity correlation, $C(r)$, for different values of the self-propulsion intensity, $f_0=v_0 \gamma$. 
The dashed black lines are the theoretical predictions, obtained by fitting the functional form given by Eq.~\eqref{eq:vv_realspace} via the function $f(x)=a\,e^{-r/\lambda}/r^{1/2}$, where $\lambda$ is given by Eq.\eqref{eq:lambda} and $a$ is a positive fitting parameter.
Panel (b): Correlation length, $\lambda$, of $C(r)$  as a function of $f_0$. 
The value of $\lambda$ has been obtained fitting the function $f(r)=a\,e^{-r/c}/r^{1/2}$, fitting also the constant $c$ to be compared with $\lambda$.
The dashed black line has been obtained evaluating Eq.~\eqref{eq:lambda} with the set of parameters of the simulation, where $\bar{x}$ has been measured numerically and reads $\bar{x}= 0.91$.
The simulations correspond to $\gamma/m=10^2$, $\epsilon=10^2$, $T=10^{-1}$, $D_r=10$.
}
\end{figure}
%---------------------------------------------------------------------

In ABP systems, the degree of  activity is often accounted for by a single dimensionless parameter, the so-called P\'eclet number, $Pe\propto v_0/D_r$, so that a decrease of $D_r$ has the same effect as an increase of $v_0$.
Actually, most of the studies concerning systems of interacting ABP are obtained via this procedure and the ABP phase diagram is usually described in terms of two parameters, density and P\'eclet number~\cite{stenhammar2014phase, digregorio2018full, costanzo2014motility, mandal2019motility, rodriguez2020phase}.

Hereafter, we demonstrate that variations of $v_0$ and $1/D_r$ are not interchangeable, as far as the spontaneous velocity alignment is 
concerned. We show that a single parameter, the P\'eclet number, is unable to fully capture the non-equilibrium dynamical properties of active particles.
In a previous study about the dense phases of overdamped ABP~\cite{caprini2020hidden}, the role of $\tau$ at fixed self-propulsion was investigated numerically and the results  were found in agreement with the theoretical predictions ($\lambda\propto \tau^{1/2}$).
In Fig.~\ref{fig:corr_self} (a), we study the velocity correlation function varying the self-propulsion intensity, $f_0$ (and, thus, $v_0$) and keeping fixed the remaining parameters. 
In Fig.~\ref{fig:corr_self}~(b),
we display the correlation length, $\lambda$,  measured fitting the functional form reported in Eq.~\eqref{eq:lambda}.
This procedure reveals that $C(r)$ is not affected by the increase of $f_0$  for a broad range of $f_0$ values for which the system remains in solid-like configurations. The correlation length (and, thus, the size of the velocity domains) remains constant.
When $f_0$ exceeds a threshold value (for $f_0 > 5\times 10^3$), the function $C(r)$  decays faster just because a solid-hexatic transition takes place.
The faster decay, corresponding to a decrease of the correlation length,
is not surprising since the lack of orientational order in the hexatic phase and periodic order in the liquid phase has been recognized as one of the main reasons for the $\lambda$ decrease~\cite{caprini2020hidden}.
As discussed in the literature (see for instance Ref.~\cite{bialke2012crystallization, digregorio2018full, caprini2020hidden}), the
occurrence of positional order is mainly controlled by the P\'eclet number and, thus, by the increase of $v_0$ and the decrease of $D_r$.
Here, we argue that, to the best of our knowledge, there is no numerical quantitative validation of the symmetric action of $v_0$ and $1/D_r$ in the phase diagram of ABP  and its evidence is at most qualitative.
In other words, it is not clear if by changing $Pe$ through $v_0$ or $1/D_r$  one could shift the transition lines of the phase diagram.
Finally, for values of $f_0$ producing spatial inhomogeneity (namely for $f_0>2\times10^4$ corresponding to $v_0>2\times 10^2$), $\lambda$ increases again revealing a non-monotonic behavior.
This effect is due to the phase-separation inducing a local increase of the density and thus the growth of $\lambda$ in the denser phase, as already observed in Ref.~\cite{caprini2020hidden}.

We also stress that our numerical results in the solid phase are supported by the main prediction, Eq.~\eqref{eq:vv_realspace} and Eq.~\eqref{eq:lambda}.
Indeed, the correlation length, $\lambda$, does not contain an explicit dependence on $f_0$ (and, thus, $v_0$). This parameter appears as a simple prefactor in the shape of $\langle \mathbf{v}(r) \cdot \mathbf{v}(0) \rangle$, specifically, through the active temperature.
Thus, cannot deeply affect the occurrence of velocity alignment, except for values of $f_0$ comparable with $T$ as detailed shown in Sec.~\ref{sec:6}.

\section{The asymmetric role of mass and viscosity}\label{sec:5}
%------------------------ FIG.2--------------------------------------
\begin{figure*}[!t]
\centering
\includegraphics[width=0.9\linewidth,keepaspectratio]{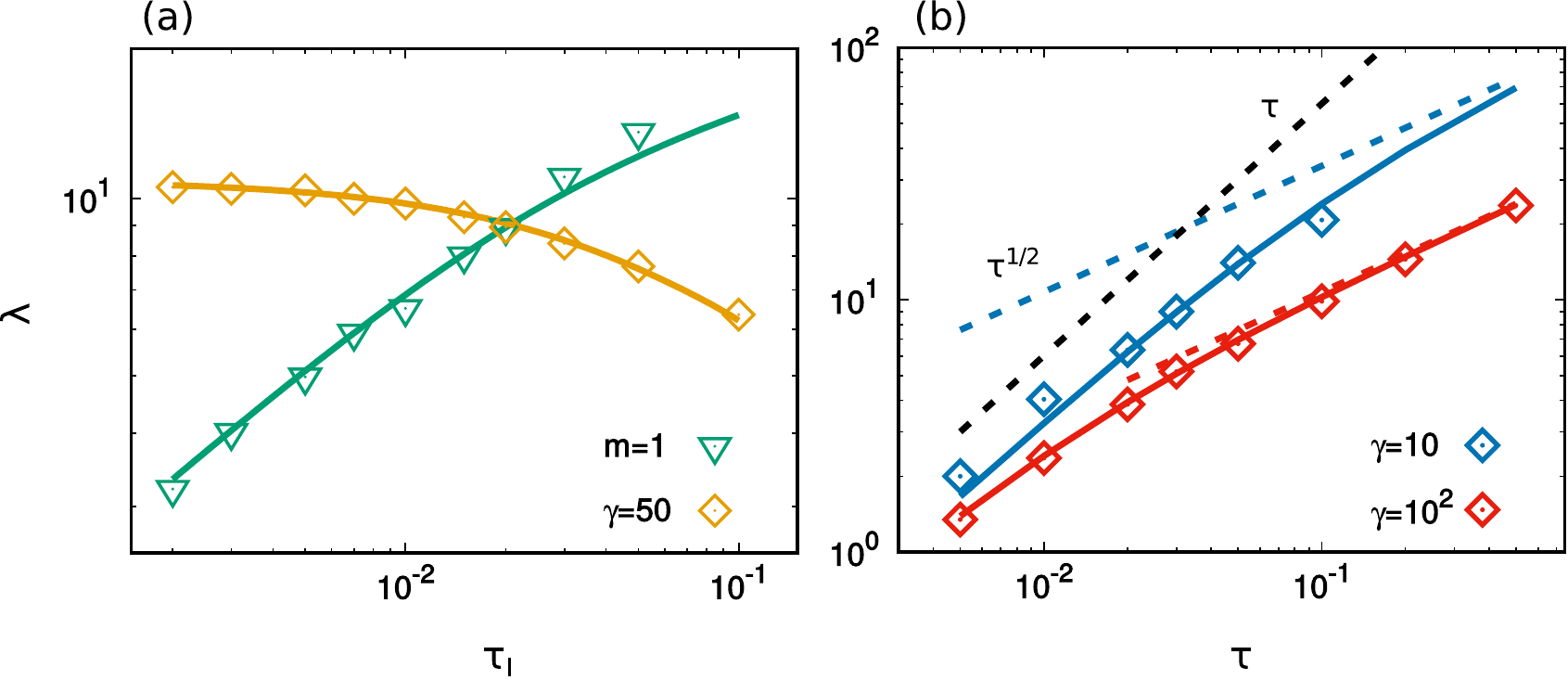}
\caption{\label{fig:under_lambda}
Correlation length for different values of  mass and viscosity. Panel~(a): Correlation length, $\lambda$, of $C(r)$ as a function of the inertial time, $\tau_I$. The green and yellow data have been obtained by varying $\gamma$ at $m=1$ and $m$ at $\gamma=50$. 
Panel~(b): $\lambda$ as a function of the persistence time, $\tau$, for two different values of $\gamma=10$ (blue points) and $10^2$ (red points). 
In both panels the points are obtained from numerical simulations while the solid lines from the theoretical prediction, Eq.~\eqref{eq:lambda}.
The dashed blue and red lines in panel~(b) are obtained from Eq.~\eqref{eq:lambda_over} and, finally, the dashed black line is an eye-guide to evidence the linear behavior with $\tau$.
The numerical values of $\lambda$ have been obtained fitting the function $f(r)=a\,e^{-r/c}/r^{1/2}$, where $c$ is the estimate of $\lambda$.
The remaining parameters of the simulations are $\epsilon=10^2$, $T=10^{-1}$ and $f=5\times 10^3$.
}
\end{figure*}
%---------------------------------------------------------------------

 In passive systems,
 the role of the inertial forces could be encapsulated in a single parameter, the inertial time, $t_I=m/\gamma$,
corresponding to  the ratio between the mass and the solvent viscosity.
Such a time controls the relaxation towards  equilibrium, 
but does not affect the steady-state properties of the system.
By contrast, as we show hereafter,  in the ABP case,  the scenario is different and reveal the non-symmetric role played by mass and inverse viscosity, and their influence on the steady-state properties of the system and on the dynamical collective phenomena reported so far.

Fig.~\ref{fig:under_lambda} (a) displays the correlation length, $\lambda$, numerically extracted from $C(r)$ for different values of $\tau_I$. The green and orange curves are obtained varying $m$ at fixed $\gamma$ and varying $\gamma$ at fixed $m$, respectively, and clearly  show different results for the same $\tau_I$ but different values of $m$ and $\gamma$. 
In particular, if $\tau_I$ is increased by varying $\gamma$, $\lambda$ reaches a constant value while, if $\tau_I$ is increased by varying $m$, $\lambda$ monotonically decreases with $m$. This is consistent with the prediction~\eqref{eq:lambda}, that in the underdamped regime where the inertial time is the larger one, $\tau_I \gg \tau$, explicitly reads:
\begin{equation}
\label{eq:lambda_under}
\lambda^2_u= \bar{x}^2 \frac{3}{4}\frac{\tau^2}{m}\left(U''(\bar{x})+\frac{U'(\bar{x})}{\bar{x}}\right)  \,.
\end{equation}
On the contrary, if $\tau_I$ is decreased by varying $\gamma$, $\lambda$ monotonically decreases while, if $\tau_I$ is decreased by varying $m$, $\lambda$ approaches a constant value, consistently with the outcome of Eq.\eqref{eq:lambda}, in the overdamped regime, i.e. when $\tau_I \ll \tau$:
\begin{equation}
\label{eq:lambda_over}
\lambda^2_o = \bar{x}^2 \frac{3}{4}\frac{\tau}{\gamma}\left(U''(\bar{x})+\frac{U'(\bar{x})}{\bar{x}}\right)  \,.
\end{equation}
This asymmetric role of mass and inverse viscosity is a pure non-equilibrium effect without a passive counterpart suggesting that a single parameter is not enough to describe the dynamical properties of far equilibrium systems.

In Fig.~\ref{fig:under_lambda} (b), we display $\lambda$ as a function of $\tau$ for two different values of $\gamma$ and $m=1$ to evaluate how inertial forces affect the scaling with the persistence time of the active force.
At first, we observe that the effect of the inertial forces is to reduce the correlation length of the spatial velocity correlation through the constant prefactor $1/(1+\tau_I/\tau)$ appearing in the expression of $\lambda^2$, Eq.~\eqref{eq:lambda}.
The comparison between the overdamped prediction, Eq.\eqref{eq:lambda_over} (dashed lines) and the numerical data (points) 
reveals a fair agreement with the prediction of Eq.~\eqref{eq:lambda} (solid lines).
The  prefactor approaches 1 in the overdamped regime, for $\tau_I \ll \tau$, giving rise to the behavior $\lambda \propto \tau^{1/2}$ that has been already reported in Ref.~\cite{caprini2020hidden}.
For $\tau \leq \tau_I$, the inertia starts playing a role in decreasing the value of $\lambda$.
For very small values of $\tau_I$, inertial effects cannot be appreciated since they could be observed only when $\tau$ 
is such that $\lambda <\sigma$ corresponding to particle velocities at different positions almost  uncorrelated. In this regime of parameters, $\lambda \propto \sqrt{\tau}$ in the whole range of $\tau$ where the velocity field has a spatial structure. 
On the contrary, Eq.~\eqref{eq:lambda_under} shows that in the regime $\tau_I \gg \tau$, the prefactor reduces to $\tau^2/m$ in such a way that $\lambda \propto \tau$.
Thus, when $\tau_I$ is large, the correlation length displays two distinct regimes with $\tau$ that are visible in Fig.~\ref{fig:under_lambda} (b), for $\gamma=10$ (red curve): a linear increase for small values of $\tau$, such that when $\tau_I \gg \tau$, is followed by the overdamped scaling, $\lambda\propto \tau^{1/2}$, always occurring in the opposite regime, $\tau_I \ll \tau$.

\section{Role of the temperature}\label{sec:6}

%------------------------ FIG.2--------------------------------------
\begin{figure}[!t]
\centering
\includegraphics[width=0.7\linewidth,keepaspectratio]{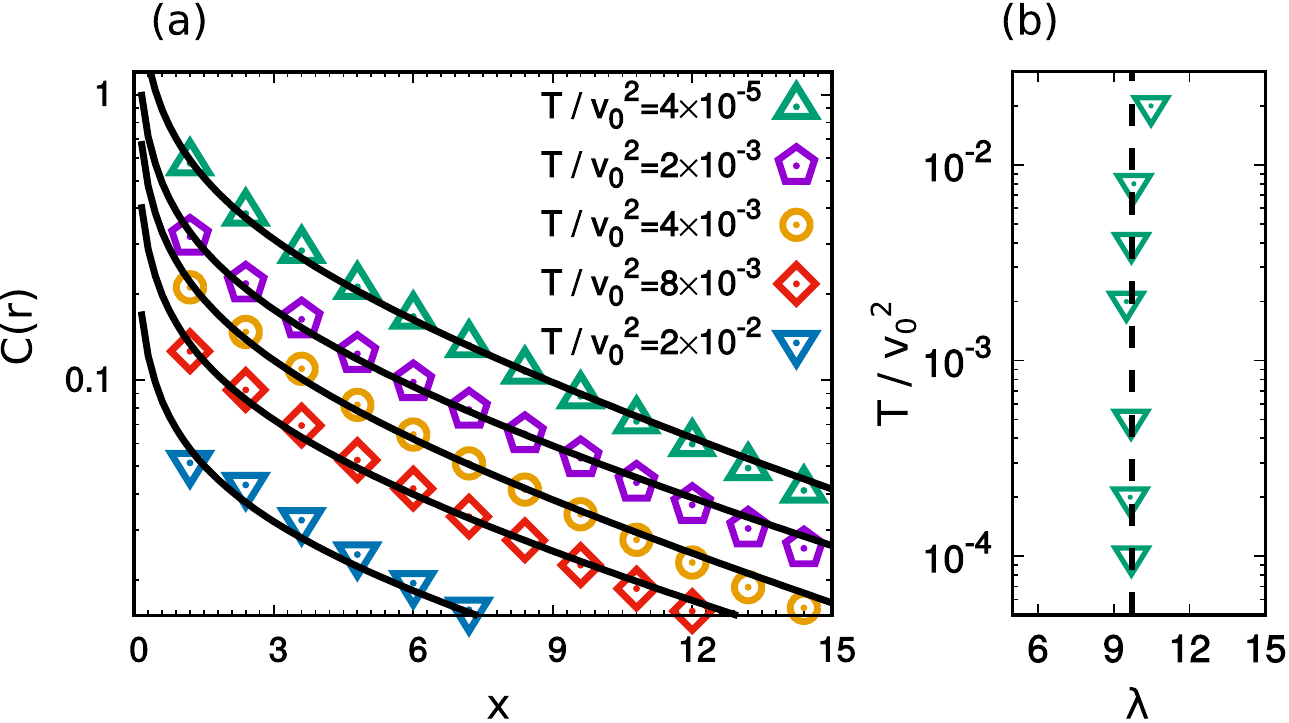}
\caption{\label{fig:Fig3}
Panel (a): Spatial velocity correlation, $C(r)$, for different values of the self-propulsion intensity, $f_0=v_0 \gamma$. 
The dashed black lines are the theoretical predictions, obtained by fitting the functional form given by Eq.~\eqref{eq:vv_realspace} via the function $f(r)=a\, e^{-r/\lambda}/r^{1/2}$, where $\lambda$ is given by Eq.~\eqref{eq:lambda} and $a$ is a fitting parameter.
Panel (b): $T/v_0^2$ vs correlation length, $\lambda$. %of $\langle \mathbf{v}(\mathbf{x}) \cdot\mathbf{v}(\mathbf{0}) \rangle / \langle \mathbf{v}^2 \rangle$ 
The values of $\lambda$ have been extracted from the data fitting the function $f(r)=a\,e^{-r/c}/r^{1/2}$, fitting also the constant $c$. 
The simulations correspond to $\gamma=10^2$, $m=1$, $\epsilon=10^2$, $v_0=50$, $D_r=10$.
}
\end{figure}
%---------------------------------------------------------------------

%{\color{red}
The temperature, $T$, is crucial 
in determining whether it is possible to detect the spontaneous velocity alignment and the occurrence of spatial velocity correlations.
Fig.~\ref{fig:Fig3} shows $C(r)$ at fixed $f_0$ and $D_r$, for different values of $T$ and keeping fixed $\gamma$ and $m$ (and, thus, $T_a$).
Interestingly, $C(r)$ decays with distance at the same rate,
but its amplitude decreases until it approaches an almost flat vanishing shape when $T$ is sufficiently large.
Our observations are in agreement with the theoretical prediction~\eqref{eq:vv_realspace}, as shown by the comparison between points and solid lines in Fig.~\ref{fig:Fig3}.
In particular, in panel~(a) the insensitivity of $\lambda$ to changes of $T$ is numerically corroborated by the comparison with the theoretical prediction. 
The solvent temperature $T$ only affects the amplitude of the normalized spatial profile of the velocity correlation entering the analytical expression for $C(r)$ just through the term $\langle \mathbf{v}^2 \rangle$ (proportional to $T_k$, Eq.~\eqref{eq:prediction_kinetictemperature2}). Indeed, its value increases when $T$ grows at variance with the expression for $\langle \mathbf{v}(r) \cdot \mathbf{v}(0) \rangle$ that remains unchanged for $r>\sigma$ (Eq.~\eqref{eq:vv_realspace}).
Hence, the amplitude of $C(r)$ for each $r>\sigma$ is controlled by the ratio $T_a/T$, through a function $\propto 1/(T/T_a + \alpha)$ where $\alpha$ is constant with respect to $T$ and $T_a$.
To summarize, a change in the solvent temperature can be mapped onto a change of the active temperature so that the relative contribution of the active and thermal fluctuations are mainly controlled by the non-dimensional ratio $T_a/T$. 
%We stress that the influence of $T_a/T$ on the dynamical collective phenomena presented so far is quite poor at variance with any equilibrium ferromagnetic scenario where the correlation length is mainly controlled by the temperature.
%

%We remark that these conclusions apply to solid-like configurations and argue that, for the broad range of temperatures in the simulations, the large values of the packing fraction considered in this study prevent the transition from solid-like to hexatic-like behavior even for passive systems (such that $f_0=0$).

We remark that these conclusions apply to solid-like configurations and argue that the transition from solid-like to hexatic-like behavior does not occur for the broad range of temperatures of the simulations neither in passive systems (such that $f_0=0$) because of the large values of the packing fraction considered in this study.

%In principle, we cannot assume that the increase of $T$ could have the same marginal influence both in hexatic and liquid configurations where, in principle, $T$ could affect also the correlation length.

Finally, some authors claimed the need to use alignment interactions to get consistent spatial structures in the velocity correlations~\cite{sarkar2020minimal} or asserted that their numerical simulations did not produce any evidence of the 
existence of velocity domains~\cite{caporusso2020motility}.
We believe that the these claims are a consequence of the range of temperatures considered in their numerical system, which were perhaps too large compared to $T_a$ according to the predictions~\eqref{eq:vv_realspace} and~\eqref{eq:prediction_kinetictemperature2}.
Other relevant causes motivating those claims could the lack of periodic order, as it occurs in homogenous active liquids.

\section{Conclusion}
\label{Sec:Conclusion}

In this article, we have studied the solid and the dense cluster regimes
of a system of interacting active particles evolving 
according to the underdamped version of the Active Brownian Particles model.
Our first target was shedding light on an emergent collective phenomenon, namely the spatial ordering of the velocity field.
This phenomenon was already observed in systems of overdamped, athermal ABP, but demanded further investigation via more realistic dynamics.
The underdamped dynamics is the natural approach to include the inertial forces and the effect of thermal noise.
We confirmed the spontaneous occurrence of velocity domains 
and quantify their average size by measuring the  correlation length of the spatial velocity correlation function.
We corroborated our numerical findings employing theoretical arguments analytically predicting both the spatial shape of the velocity correlation and the parameter dependence of its correlation length. 

We have also shown that a single dimensionless parameter, such as the P\'eclet number (usually defined as proportional to the swim velocity and to the persistence time) fails to fully describe the velocity alignment phenomenon in dense ABP systems or their phase-separated configurations.
A change in the persistence time cannot be mapped onto a change in the swim velocity, in contrast with the widespread opinion in the literature.
Indeed, the size of the domains (corresponding to the correlation length of the spatial velocity correlation) increases with the persistence time while remains constant with the self-propulsion intensity (that is proportional to the swim velocity).
Despite the P\'eclet number has been intensively used to describe the structural properties of the system (usually, the phase diagram is described as a function of density and P\'eclet number), it gives an insufficient description of the spatial properties of the velocity field.
To the best of our knowledge, a phase diagram obtained by changing the P\'eclet number through the persistence time in alternative to the swim velocity has not been yet evaluated in the case of purely repulsive ABP. 
Our analysis suggests that a three-dimensional phase diagram is needed to characterize the phenomenology of Active Brownian Particles (at least, concerning the dynamical collective phenomena) and further investigations about the MIPS transition line or the solid-hexatic and hexatic-liquid transitions could be needed.

We have also explored the role of the inertial forces finding fascinating results that hold in solid configurations or the dense clusters of MIPS.
Inertial forces introduce a typical time, $\tau_I$, in addition to the persistence time of the active force.
When the former is the larger one, the correlation length is decreased providing two main results:
I) inertia reduces the velocity alignment  with respect to the overdamped case. ii) The scaling of the correlation length with the persistence time is deeply affected.
A linear regime, $\propto \tau$, for an initial broad interval of $\tau$ values, appears before the overdamped regime, scaling as $\propto \sqrt{\tau}$, takes over as the persistence time becomes larger than the inertial time.
Last but not least, we surprisingly observe a further non-equilibrium effect manifesting in the non-symmetric role of mass and inverse viscosity in the correlation length of the spatial velocity correlations.
While the $\tau$ scaling is controlled by the inertial time, we show that the value of the correlation length explicitly depends on the mass and viscosity values separately and not only on their ratio (the inertial time).
This observation suggests further investigations to test the role of the inertia on the phase diagram varying separately both $\gamma$ and $m$ (and not just the inertial time), with particular attention to the coexistence line of the Motility Induced Phase Separation that could be deeply affected.

Finally, we highlight the non-thermal nature of the collective phenomenon described so far that is  marginally affected by a
temperature change, at variance with equilibrium models, such as the $XY$ models.
The increase of the solvent or active temperature leaves unchanged the correlation length (and, thus, the size of the velocity domains), at least in the dense configurations evaluated in this work. 
The use of $T/ T_a$ can be recognized as the dimensionless parameter necessary to compare the strengths of active force and thermal fluctuations. The increase of this ratio reduces the amplitude of the rescaled velocity correlation because of the $T/T_a$ dependence on the kinetic temperature.
We conclude that, to observe the velocity domains (or, equivalently, spatial structure in the velocity correlations) it is necessary to fix the solvent temperature rather smaller than the active temperature so that the active force term (that produces effective alignment interactions) is not overwhelmed by the uncorrelated thermal fluctuations.

\appendix

\section{Overdamped ABP dynamics}\label{app:1}
In this Appendix, we report the numerical details employed to simulate overdamped ABP to measure the spatial velocity correlations in Fig.~\ref{fig:corr_self}.
Each particle is described by an equation of motion for its position $\mathbf{x}_i$:
\begin{equation}
\label{eq:motion_over}
\gamma\dot{\mathbf{x}}_i = \mathbf{F}_i + \mathbf{f}^a_i + \sqrt{2 \gamma m T} \boldsymbol{\eta}_i \,,
\end{equation}
where the parameters $\gamma$, $T$, $m$ have the same physical meaning as in Eq.~\eqref{eq:motion}. 

The term $\boldsymbol{\eta}$ is a white noise vector with zero average and unit variance due to the collision by the solvent particles.
The force term $\mathbf{F}_i$ models steric interactions between particles and is derived from the same potential used to simulate Eq.~\eqref{eq:motion}.
Finally, $\mathbf{f}^a_i$ represents the active force, that in the literature based on ABP simulations is usually expressed as 
$$
\mathbf{f}^a_i=\gamma v_0 \mathbf{n}_i \,.
$$
This is consistent with our notation and, in particular, with the swim velocity definition, Eq.~\eqref{eq:def_swimvelocity}.
In this system, the velocity vector employed to calculate the spatial velocity correlation function is obtained from the the relation ${\mathbf{v}}_i=\dot{\mathbf{x}}_i$.% that by definition corresponds to the velocity also in overdamped dynamics.

\section{Derivation of Eq.~\eqref{eq:vv_fourierspace}}\label{app:2}

In order to obtain the velocity correlation function in the Fourier space, i.e. Eq.~\eqref{eq:vv_fourierspace}, 
we shall make two simplifying assumptions  in Eq.~\eqref{eq:motion}:
\begin{itemize}
\item[i)] We consider the AOUP model, assuming that $\mathbf{f}^a_i$ is given by Eq.~\eqref{eq:motion2}.
\item[ii)] Each particle performs small oscillations around a node of a hexagonal lattice so that the 
total inter-particle potential can be approximated as the sum of quadratic terms.
\end{itemize}
Introducing the displacement $\bu_i$ of the particle $i$ with respect to its equilibrium position, $\bRz_i$, namely
\begin{equation*}
\bu_i=\bR_i-\bRz_i \,,
\end{equation*}
the pair potential, in the harmonic approximation, reads:
$$
U_{tot}\approx m \frac{\omega_E^2 }{2}\sum_{i\neq j} (\bu_j-\bu_i)^2 \,,
$$ 
where 
$$
\omega_E^2 = \frac{1}{2 m}\left(U''(\bar x) + \frac{U'(\bar x)}{\bar x} \right) \, .
$$
Therefore, the equations of motion become:
\begin{subequations}
\label{dynamicequation2}
\begin{align}
%\begin{eqnarray}
%&&
\dot{\mathbf{f}}^a_i&=-\frac{1}{\tau}\mathbf{f}^a_i + \sqrt{\frac{2 } { \tau}}  f_0 \boldsymbol{\xi}_i \,,\\ 
%&&
\dot{\mathbf{v}}_i(t) &=  -\frac{\gamma}{m} \mathbf{v}_i(t) 
+\omega_E^2 \sum_j^{n.n} (\bu_j-\bu_i) + \frac{\mathbf{f}^a_i}{m} + \sqrt{2 \frac{\gamma}{ m }T} \boldsymbol{\eta}_i \,. %\nonumber
 %\label{dynamicequation2}
%\end{eqnarray}
\end{align}
\end{subequations}
where the sum is over nearest neighbour only.
%where $K$ is the strength of the potential in the harmonic approximation, i.e. $U\approx\frac{K}{2} (\bu_j-\bu_i)^2$, which reads
Introducing the discrete Fourier transforms of the displacement about the equilibrium positions,  velocity  and  active force 
$\hat{ \mathbf{u}}_q,\hat{ \mathbf{v}}_q,\hat{ \mathbf{f}}^a_q $, respectively,
the equations of motion~\eqref{dynamicequation2} can be written in the Fourier Space:
\begin{subequations}
\label{eq:app_qdyn}
\begin{align}
&\frac{d}{dt}
\hat{\mathbf{v}}(\mathbf{q})
=-\frac{\gamma}{m} \hat{\mathbf{v}}(\mathbf{q})- \omega^2(\mathbf{q}) \hat{\mathbf{u}}
(\mathbf{q})+ \frac{\hat{\mathbf{f}    }^a(\mathbf{q})}
{m}
+ \sqrt{2 \gamma \frac{T}{m}} \,\hat{\boldsymbol{\eta}}(\mathbf{q})\\
&\tau \frac{d}{dt}
 \hat{ \mathbf{f}}^a(\mathbf{q}) = -\hat{ \mathbf{f}}^a(\mathbf{q})  + f_0\sqrt{2 \tau} \hat{\boldsymbol{\xi}}(\mathbf{q})  \,,
\end{align}
\end{subequations}
where $\mathbf{q}=(q_x, q_y)$ are the Cartesian components of vectors of the reciprocal Bravais lattice. 
The frequency $\omega^2(\mathbf{q})$ reads:
\begin{flalign}
%\begin{aligned}
\label{eq:app_omegaq}
\omega^2(\mathbf{q})
&=-2 \omega_E^2 \Bigl[\cos(q_x \bar x) +2\cos\Bigl(\frac{1} {2} q_x \bar x\Bigr)\cos \Bigl(\frac{\sqrt 3} {2} q_y \bar x\Bigr)-3\Bigr] \nonumber\\
& \approx \frac{3}{2} \omega_E^2 \bar x^2 \mathbf{q}^2 + O(\mathbf{q}^4)\,,
%\end{aligned}
\end{flalign}
where in the last line we have performed a Taylor expansion around $\mathbf{q}=0$.
Solving the dynamics~\eqref{eq:app_qdyn}, we get the final expression for the positional correlation function:
\begin{equation*}
\label{eq:u_fourierspacebis}
\langle \hat{\mathbf{u}}(\mathbf{q})\cdot\hat{ \mathbf{u}}(-\mathbf{q})\rangle = \frac{2T}{ m \omega^2(\mathbf{q})}  +\frac{2 f_0^2 }{m  \omega^2(\mathbf{q})}\frac{\tau}{\gamma} \frac{1}{1+\frac{\tau^2}{1+\tau/\tau_I}\omega^2(\mathbf{q})}
\end{equation*}
and the velocity correlation functions in the Fourier space:
\begin{equation}
\label{eq:v_fourierspacebis}
\langle \hat{\mathbf{v}}(\mathbf{q})\cdot \hat{\mathbf{v}}(-\mathbf{q})\rangle = \frac{2T}{ m}  +\frac{2 f_0^2 }{m}\frac{\tau}{\gamma}\frac{1}{1+\tau/\tau_I} \frac{1}{1+\frac{\tau^2}{1+\tau/\tau_I}\omega^2(\mathbf{q})}
\end{equation}
Equation~\eqref{eq:v_fourierspacebis} coincides with Eq.~\eqref{eq:vv_fourierspace} using the definition of $T_a$.

%%%%%%%%%%%%%%%%%%%%%%%%%%%%%%%

\section{Derivation of Eq.\eqref{eq:vv_realspace}}\label{app:3}

The velocity real space correlation, i.e. Eq.~\eqref{eq:vv_realspace}, is obtained by inverting formula~\eqref{eq:v_fourierspacebis}:
\begin{equation}
\label{eq:vv_fourierspace2r}
\langle \vv_{\bf x} \cdot \vv_{\bf x'}\rangle
 = \frac{2T}{ m} \delta_{{\bf x},{\bf x}'} +
\frac{2 f_0^2 }{m}\frac{\tau}{\gamma}\frac{1}{1+\tau/\tau_I} 
% \nonumber\\&&
  \sum_{\bq} 
 \frac{e^{i\bq (\bx-\bx')}}{1+\frac{\tau^2}{1+\tau/\tau_I} \omega^2(\mathbf{q})}
\end{equation}
For large particle separations, $r=| {\bf x} - {\bf x'}|>\sigma$, the first term is negligible while the second term can be evaluated by performing the following approximations:
i) the finite lattice sum is replaced by a double integral over  $(q_x, q_y)$ variables,
ii)  the frequency is replaced by its small $\mathbf{q}$-expansion and
iii) the limits of integration are extended from $-\infty$ to $\infty$.
Using these approximations, we have:
   \begin{equation*} % requires amsmath; align* for no eq. number
\langle \vv_{\bf x} \cdot \vv_{\bf x'}\rangle  \approx   2\frac{f_0^2 \tau}{m \gamma}
 \frac{1}{2\pi} \frac{\bar x^2}{\lambda^2}  \frac{1}{1+\frac{\tau}{\tau_I} }  K_0(r/\lambda)  \,,
   \end{equation*}
   where the coherence length (or correlation length) $\lambda$ is given by:
    \begin{equation}
\label{eq:app_lambda}
      \lambda^2  \equiv  \frac{3}{2}\bar x^2 \frac{ \omega_E^2\tau^2}{1+\frac{\tau}{\tau_I}}  
   \end{equation}
and $K_0(r/\lambda)$ is the zero-order modified Bessel function of the second kind which has the following asymptotic behavior when $r/\lambda \gg1$:
$$
K_0(r/\lambda) \approx \Bigl(\frac{\pi \lambda}{2  r}\Bigr)^{1/2}  e^{-  r/\lambda} \,.
$$
Therefore, for large separations, we find the following approximation:
   \begin{equation} 
   \langle \vv_{\bf x} \cdot \vv_{\bf x'}\rangle
   \approx 2\frac{f_0^2 \tau}{m \gamma}\frac{1}{1+\frac{\tau}{\tau_I} }
  \frac{\bar x^2}{\lambda^2}  \left( \frac{ \lambda}{8 \pi r}\right)^{1/2} e^{-r/\lambda} \,.
   \end{equation}
Switching to a continuous notation such that $\vv_{\bf x} \to \mathbf{v}(\mathbf{x})$, fixing $r=|\mathbf{x}-\mathbf{x}'|$ and formally dividing by the velocity variance $\langle \vv_{\bf x}^2\rangle$, we obtain Eq.~\eqref{eq:vv_realspace} after using the definition of $T_a$ while Eq.~\eqref{eq:app_lambda} coincides with Eq.~\eqref{eq:lambda}.

\section{Kinetic temperature: Eq.~\eqref{eq:prediction_kinetictemperature2}}\label{app:4} %Variance

To obtain an analytical expression for the kinetic temperature, we need to calculate Eq.~\eqref{eq:vv_fourierspace2r} in $r=|\mathbf{x}-\mathbf{x}'|=0$. 
In this case, we need to consider the exact expression of $\omega^2(\mathbf{q})$ without employing any small $\mathbf{q}$ expansion.  
We replace the sum by a double integral over a finite domain:
 \begin{equation}
 \langle \vv_{\bf x} \cdot \vv_{\bf x}\rangle  =\frac{2 T}{m}+ \nonumber
 \frac{2 f_0^2 }{m}\frac{\tau}{\gamma} \frac{1}{1+\tau/\tau_I+6 \omega_E^2\tau^2}
 \int_{-\pi}^\pi \frac{dk_1}{2\pi} \int_{-\pi}^\pi \frac{d k_2}{2\pi}
 \frac{1}{1-z s(k_1,k_2)}
 \label{eq:greenfunction}
 \end{equation}
 where we have performed a change of variables of integration and introduced
 $s(k_1,k_2)$, the so-called  structure function of the triangular lattice~\cite{guttmann2010lattice}:
 $$
 s(k_1,k_2)=\frac{1}{3}(\cos(k_1)+\cos(k_2)+\cos(k_1+k_2)) \,,
 $$
 with
  $$
 z=\frac{1}{1+\frac{1+\tau/\tau_I}{6 \omega_E^2 \tau^2}} \,.
$$
 In detail, one can evaluate the  integral as:
\begin{equation*}
\int_{-\pi}^\pi \frac{dk_1}{2\pi} \int_{-\pi}^\pi \frac{d k_2}{2\pi}
 \frac{1}{1-z s(k_1,k_2)} 
=\frac{6 }{\pi z \sqrt{c}} {\bf K}(k) \,,\\  
\end{equation*}
where ${\bf K}(k)$ is the complete elliptic integral of the first kind:
$$
{\bf K}(k)=\int_0^{\pi/2} \frac{ d\theta}{\sqrt{1-k^2 \sin^2(\theta)}} \,,
$$
with
\begin{eqnarray*}
&&
c=\frac{9}{z^2}-3+\sqrt{3+\frac{6}{z}}\\&&
k=2\frac{(3+\frac{6}{z})^{1/4}}{c^{1/2}} \,.
\end{eqnarray*}
Hence, in the limit $\tau\to \infty$, we have $z\to 1$ and the integral in Eq.~\eqref{eq:greenfunction} weakly (in fact, logarithmically)
diverges for any two-dimensional lattice, being connected to the fact that
the probability of returning to the origin by a random walker in two dimensions is certain.
However, in the same limit, the dependence on $\tau$ of the prefactor in front of the integral makes the resulting contribution of the self-propulsion to the velocity variance vanishingly small. This can be seen as a consequence of the well-known fact that the velocity of active particles also depends on the forces they experience in such a way that they are slower in those regions where the curvature of the local potential is high.
Finally, upon defining:
$$
\mathcal{I} = \frac{6 }{z \sqrt{c}} {\bf K}(k)  \,,
$$
we get the exact expression for the kinetic temperature reported in Eq.~\eqref{eq:vv_realspace}.
Because of the definitions of $k$, $c$ and $z$, the term $\mathcal{I}$ depends only on $\tau$, $\tau_I$ and $\omega_E$ and, thus, is independent of $T$ and $T_a$.
%%%%%%%%%%%%%%%%%%%%%%%%%%%%%%%%%%%%%%%%%%

%%%%%%%%%%%%%%%%%%%%%%%%%%%%%%%%%%%%%%%%%%
\acknowledgments{LC and UMBM acknowledge support from the MIUR PRIN 2017 project 201798CZLJ. In addition, LC and UMBM warmly thank Andrea Puglisi for letting us use the computer facilities of his group and for discussions regarding 
some aspects of this research.}

%\reftitle{References}


%\externalbibliography{yes}
%\bibliography{your_external_BibTeX_file}
%\externalbibliography{yes}

\bibliography{under}

%=====================================
% References, variant B: internal bibliography
%=====================================
%\begin{thebibliography}{999}
% Reference 1
%\bibitem[Author1(year)]{ref-journal}
%Author1, T. The title of the cited article. {\em Journal Abbreviation} {\bf 2008}, {\em 10}, 142--149.
% Reference 2
%\bibitem[Author2(year)]{ref-book}
%Author2, L. The title of the cited contribution. In {\em The Book Title}; Editor1, F., Editor2, A., Eds.; Publishing House: City, Country, 2007; pp. 32--58.
%\end{thebibliography}

% The following MDPI journals use author-date citation: Arts, Econometrics, Economies, Genealogy, Humanities, IJFS, JRFM, Laws, Religions, Risks, Social Sciences. For those journals, please follow the formatting guidelines on http://www.mdpi.com/authors/references
% To cite two works by the same author: \citeauthor{ref-journal-1a} (\citeyear{ref-journal-1a}, \citeyear{ref-journal-1b}). This produces: Whittaker (1967, 1975)
% To cite two works by the same author with specific pages: \citeauthor{ref-journal-3a} (\citeyear{ref-journal-3a}, p. 328; \citeyear{ref-journal-3b}, p.475). This produces: Wong (1999, p. 328; 2000, p. 475)

%%%%%%%%%%%%%%%%%%%%%%%%%%%%%%%%%%%%%%%%%%
%% Optional
%\sampleavailability{Samples of the compounds ...... are available from the authors.}

%% for journal Sci
%\reviewreports{\\
%Reviewer 1 comments and authors’ response\\
%Reviewer 2 comments and authors’ response\\
%Reviewer 3 comments and authors’ response
%}

%%%%%%%%%%%%%%%%%%%%%%%%%%%%%%%%%%%%%%%%%%
\end{document}